\documentclass[english,aps,prl,superscriptaddress,twocolumn,address,showacknowledgments,showpacs]{revtex4}
\usepackage[T1]{fontenc}
\usepackage[latin9]{inputenc}
\usepackage{amstext}
\usepackage{amsmath}
\usepackage{amssymb}
\usepackage{graphicx}
\usepackage{esint}

\makeatletter
\@ifundefined{textcolor}{}
{%
 \definecolor{BLACK}{gray}{0}
 \definecolor{WHITE}{gray}{1}
 \definecolor{RED}{rgb}{1,0,0}
 \definecolor{GREEN}{rgb}{0,1,0}
 \definecolor{BLUE}{rgb}{0,0,1}
 \definecolor{CYAN}{cmyk}{1,0,0,0}
 \definecolor{MAGENTA}{cmyk}{0,1,0,0}
 \definecolor{YELLOW}{cmyk}{0,0,1,0}
}

\makeatother

\usepackage{babel}
\begin{document}

\title{Optomechanical Metamaterials: Dirac polaritons, Gauge fields, and
Instabilities}

\author{M. Schmidt}

\address{University of Erlangen-N\"urnberg, Staudtstr. 7, Institute for Theoretical
Physics, D-91058 Erlangen, Germany}

\author{V. Peano}

\address{University of Erlangen-N\"urnberg, Staudtstr. 7, Institute for Theoretical
Physics, D-91058 Erlangen, Germany}

\author{F. Marquardt}

\address{University of Erlangen-N\"urnberg, Staudtstr. 7, Institute for Theoretical
Physics, D-91058 Erlangen, Germany}

\address{Max Planck Institute for the Science of Light, G\"unther-Scharowsky-Stra\ss e
1/Bau 24, D-91058 Erlangen, Germany}
\begin{abstract}
Freestanding photonic crystals can be used to trap both light and
mechanical vibrations. These \textquotedbl{}optomechanical crystal\textquotedbl{}
structures have already been experimentally demonstrated to yield
strong coupling between a photon mode and a phonon mode, co-localized
at a single defect site. Future devices may feature a regular superlattice
of such defects, turning them into \textquotedbl{}optomechanical arrays\textquotedbl{}.
In this letter we predict that tailoring the optomechanical band structure
of such arrays can be used to implement Dirac physics of photons and
phonons, to create a photonic gauge field via mechanical vibrations,
and to observe a novel optomechanical instability. 
\end{abstract}


\maketitle
Studies of light interacting with nanomechanical motion have progressed
rapidly in recent years. Current successes in the field of cavity
optomechanics include the radiative cooling of a nanomechanical mode
to the quantum ground state \cite{2011_Chan_LaserCoolingNanomechOscillator,2011_Teufel_SidebandCooling_Nature},
strong coupling physics \cite{2009_Groblacher_StrongCoupling_Nature},
state transfer \cite{FioreWang2011,Verhagen2012},  radiation-mechanics
entanglement \cite{2013_Palomakilehnertentanglement}, and many more
(for a review see \cite{Aspelmeyer2013RMPArxiv}). Recently, a new
frontier is opening up: first steps have been taken towards exploring
setups with more optical and vibrational modes, e.g. by coupling two
mechanical or optical modes to investigate issues such as synchronization
\cite{Zhang2012Sync,Bagheri2013}, Brillouin cooling \cite{Bahl2012},
 phonon lasing \cite{Grudinin2010}, or wavelength conversion \cite{Hill2012WavelengthConversion,Dong2012}.
A setup particularly well suited for this avenue consists in so-called
optomechanical crystals \cite{Eichenfield2009,Safavi-Naeini2010,Safavi-Naeini2010APL,Gavartin2011PRL_OMC,2011_Chan_LaserCoolingNanomechOscillator,Safavi-Naeini2013Arxiv}.
This platform offers excellent scalability and design flexibility
in the creation of defects comprising strongly interacting co-localized
optical and mechanical modes \cite{Safavi-Naeini2010}. A superarray
of such defects (an ``optomechanical array'') is the next logical
step forward in this development. First theoretical studies have indicated
that these arrays could exhibit functionalities such as slow light
\cite{Chang2011}, quantum information processing \cite{Schmidt2012},
synchronization \cite{Heinrich2011CollDyn,2012_Holmes_Synchronization,2013_Ludwig}
and quantum many-body physics \cite{Bhattacharya2008,Tombadin2012,Xuereb2012,2012_AkramMultimodephotonphononentanglement,2013_Ludwig}.
In the present letter, we predict novel features that can be obtained
by engineering the optomechanical band structure of these arrays,
creating, in effect, optomechanical metamaterials with tailored properties. 

\begin{figure}
\includegraphics[width=0.9\columnwidth]{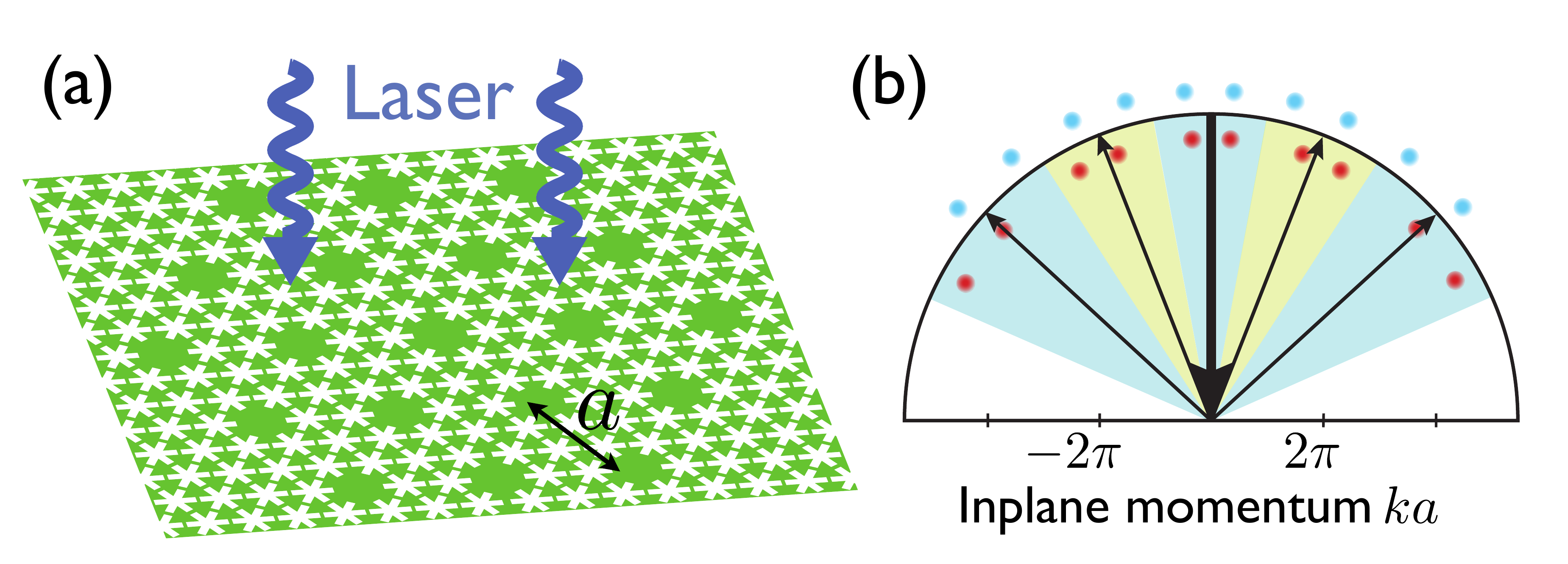}

\caption{\label{fig:Setup}Two-dimensional Optomechanical Array: (a) A patterned
optomechanical crystal slab supports localized photonic and phononic
defect modes with on-site optomechanical interaction. The defects
form a 2D superlattice, in which neighboring sites couple via photon
and phonon tunneling, and the optical modes are driven by a single
laser beam with frequency $\omega_{L}$. (b) Light will be back-reflected
from the structure into various momenta whose projections onto the
plane differ by reciprocal lattice vectors.  The arrows indicate
elastic scattering of the pump light. In addition, around each elastic
scattering peak, there will be inelastically scattered signals due
to the optomechanical interaction (schematically indicated by the
blue and red dots). These reveal the optomechanical band structure,
as discussed in the main text.}
\end{figure}

We consider a 1D or 2D lattice of identical optomechanical cells,
comprising a localized photon mode $\hat{a}_{j}$ and a phonon mode
$\hat{b}_{j}$:

\begin{equation}
\hat{H}_{j}/\hbar=\Omega\hat{b}_{j}^{\dagger}\hat{b}_{j}-\Delta\hat{a}_{j}^{\dagger}\hat{a}_{j}+\alpha_{L}(\hat{a}_{j}^{\dagger}+\hat{a}_{j})-g_{0}(\hat{b}_{j}^{\dagger}+\hat{b}_{j})\hat{a}_{j}^{\dagger}\hat{a}_{j}.
\end{equation}
 Here, $\Omega$ is the phonon mode eigenfrequency, and a laser of
amplitude $\alpha_{L}$ and detuning $\Delta\equiv\omega_{L}-\omega_{{\rm phot}}$
drives uniformly all the cells. The overlap between the evanescent
tails of the localized modes leads to tunneling of photons and phonons,
which we describe in a tight-binding model (see also \cite{Heinrich2011CollDyn,2013_Ludwig,Duan2012}).
This is similar in spirit to the Hubbard model (or, even more closely,
the Holstein model), but for photons and phonons. Our formulas will
be general, but in concrete examples we will assume nearest-neighbor
hopping $\hat{H}_{\mathbf{{\rm hop}}}=-\hbar\sum_{\langle ij\rangle}J\hat{a}_{j}^{\dagger}\hat{a}_{i}+K\hat{b}_{j}^{\dagger}\hat{b}_{i}+{\rm h.c.}$.
Then the full Hamiltonian reads $\hat{H}=\hat{H}_{S}+\hat{H}_{diss}$,
where $\hat{H}_{S}=\sum_{j}\hat{H}_{j}+\hat{H}_{{\rm hop}}$ and $\hat{H}_{{\rm diss}}$
describes the coupling to the environment, including photon (phonon)
decay at rate $\kappa$ ($\Gamma$).

For laser driving below a finite threshold (see below), the light
amplitudes $\alpha=\langle\hat{a}_{j}\rangle$ and mechanical displacements
$\beta=\langle\hat{b}_{j}\rangle$ at all cells reach a uniform steady
state, which leads to a shifted detuning $\tilde{\Delta}$ (see Supplemental
Material \cite{SMoptoarray}). We linearize the dynamics around the
steady solution, which is an excellent approximation \cite{Aspelmeyer2013RMPArxiv}
for current values of $g_{0}$. In a plane-wave basis one finds \cite{SMoptoarray}
\begin{equation}
\hat{H}_{S}\approx\hbar\sum_{k}\Omega_{k}\hat{b}_{k}^{\dagger}\hat{b}_{k}-\Delta_{k}\hat{a}_{k}^{\dagger}\hat{a}_{k}-g(\hat{a}_{k}^{\dagger}+\hat{a}_{-k})(\hat{b}_{-k}^{\dagger}+\hat{b}_{k}).\label{eq:Hamiltonian}
\end{equation}
 Here, $\hat{a}_{k}$ and $\hat{b}_{k}$ are the photonic and phononic
Bloch modes, where $\hat{a}_{k}=N^{-1/2}\sum_{j}e^{-ik\cdot r_{j}}(\hat{a}_{j}-\alpha$)
for a Bravais lattice of $N$ sites (likewise for $\hat{b}_{k}$).
We have introduced the band structure of the free phonons $\Omega_{k}=\Omega-Kf(k)$
(with $f(k)=2\cos ka$ for a 1D chain of lattice constant $a$) and
the photonic dispersion $\Delta_{k}=\tilde{\Delta}+Jf(k)$. The coupling
constant $g=g_{0}\alpha$ of the linearized interaction is enhanced
by the light amplitude $\alpha$. It comprises beamsplitter terms
$\hat{a}_{k}^{\dagger}\hat{b}_{k}$ involving Bloch modes with equal
quasimomentum and two-mode squeezing terms $\hat{a}_{k}^{\dagger}\hat{b}_{-k}^{\dagger}$
involving modes with opposite quasimomentum. 

Both fluctuations and dissipation can be included via the input-output
formalism, which yields the following quantum Langevin equations:

\begin{eqnarray}
\dot{\hat{a}}_{k} & = & (i\Delta_{k}-\kappa/2)\hat{a}_{k}+ig(\hat{b}_{k}+\hat{b}_{-k}^{\dagger})+\hat{\xi}_{k}\nonumber \\
\dot{\hat{b}}_{k} & = & (-i\Omega_{k}-\Gamma/2)\hat{b}_{k}+ig(\hat{a}_{-k}^{\dagger}+\hat{a}_{k})+\hat{\eta}_{k}\,,\label{eq:linearizedeqofmotion-2}
\end{eqnarray}
where $\hat{\xi}_{k}$ and $\hat{\eta}_{k}$ are the noise operators,
see Supplemental Material \cite{SMoptoarray}. 

The optomechanical band structure can now be studied by finding the
(complex) eigenfrequencies $\omega(k)$ for the homogeneous part of
equations~(\ref{eq:linearizedeqofmotion-2}). These are the roots
of the quartic polynomial 
\begin{equation}
\left[\left(\omega+i\kappa/2\right)^{2}-\Delta_{k}^{2}\right]\left[\left(\omega+i\Gamma/2\right)^{2}-\Omega_{k}^{2}\right]+4\Delta_{k}\Omega_{k}g^{2}=0.\label{eq:generalbandstructure}
\end{equation}
 They are complex numbers, where $-2\text{Im}\omega$ represents the
intensity decay rate of the corresponding excitation. In the noninteracting
case $g=0$, the four bands describe photon (phonon) particles and
holes with eigenfrequencies $\mp\Delta_{k}-i\kappa/2$ ($\pm\Omega_{k}-i\Gamma/2$),
respectively. The phonon bands are flat compared to the photon bands,
as for typical parameters $K\ll\Omega,J$. 

\begin{figure}
\includegraphics[width=1\columnwidth]{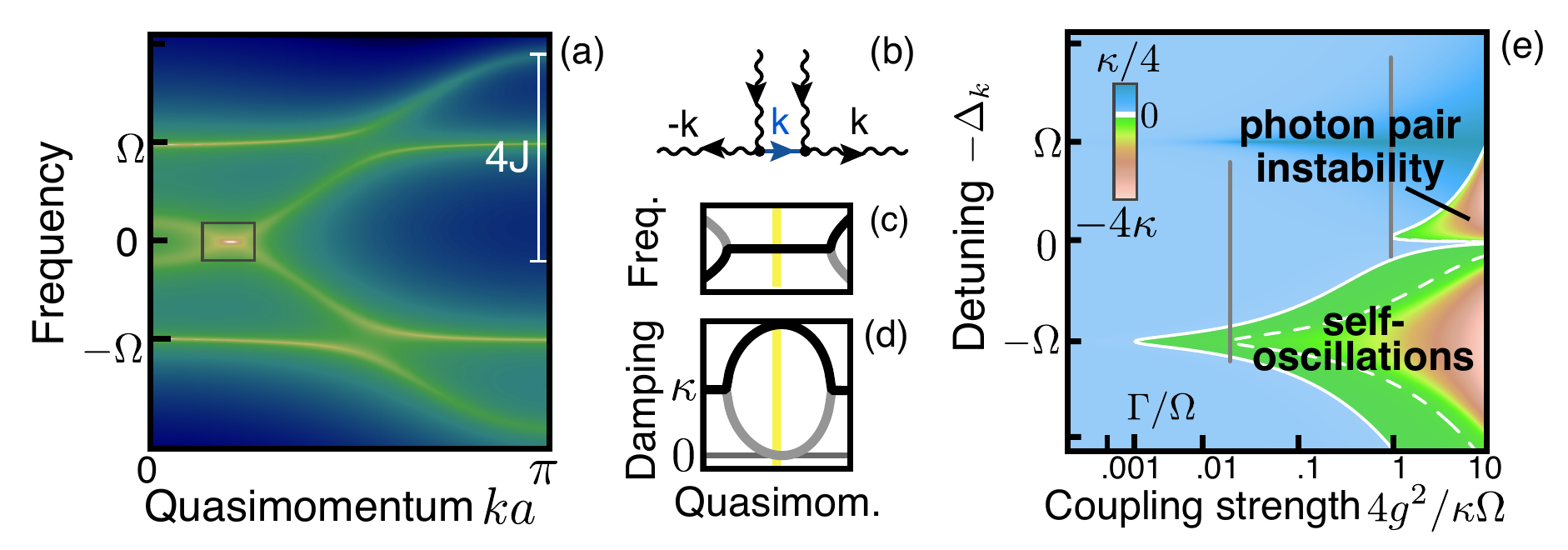}

\caption{\label{fig:Optomechanical-band-structure}Optomechanically-induced
optical parametric oscillator: (a) Photon emission spectrum $S(k,\omega)$
(see Supplemental Material \cite{SMoptoarray}) revealing the optomechanical
band structure {[}yellow corresponds to large powers{]}. The gray
box indicates the region where two photons from the drive are resonantly
converted into a photon pair with opposite quasimomenta, in a four-wave
mixing process mediated by a virtual phonon; see Feynman diagram (b).
(c-d): The band structure in this region is shown in panels (c-d),
with a distinctive behaviour in the interval $\Delta_{k}\in]-2\tilde{g},0[$.
(e) Stability diagram for the linearized dynamics in an optomechanical
array (for $\kappa=0.1\Omega$). Color: smallest damping rate as a
function of $\Delta_{k}$ and the laser-enhanced coupling strength
$g^{2}$. The zero contours delineate regions of instability (solid:
for $\Gamma/\Omega=0.002$, dashed: for $\Gamma/\Omega=0.02$). The
range of $\Delta_{k}$ plotted in Fig. \ref{fig:Optomechanical-band-structure}(a)
is indicated via the gray line to the right {[}line to the left: parameters
for Fig. \ref{fig:bandstructure-bluedet}(b-c){]}. $ $ The regions
of the photon pair creation instability (optomechanical optical parametric
oscillator) and the photon-phonon pair creation instability (optomechanical
self-oscillations, see Fig.~\ref{fig:bandstructure-bluedet}) are
indicated.}
\end{figure}

For $|\Delta_{k}|\ll\Omega$, the mechanical oscillations are faster
than the fluctuations of the field amplitude and can be adiabatically
eliminated, leading to a two-mode squeezing Hamiltonian 
\begin{equation}
\tilde{H}_{k}/\hbar=-(\Delta_{k}+\tilde{g})(\hat{a}_{k}^{\dagger}\hat{a}_{k}+\hat{a}_{-k}^{\dagger}\hat{a}_{-k})-\tilde{g}(\hat{a}_{k}^{\dagger}\hat{a}_{-k}^{\dagger}+c.c.),
\end{equation}
with $\tilde{g}=2g^{2}/\Omega$. Physically, two drive photons (arriving
perpendicularly to the structure, i.e. at in-plane quasimomentum $k=0$)
are converted, via mechanically-mediated four-wave mixing, into a
pair of counter-propagating Bloch photons in the array, see sketch
in Fig.~\ref{fig:Optomechanical-band-structure}(a). Thus, in a 2D
array, there will be entanglement in the momenta. This process then
gives rise to an instability (for $4g^{2}/\kappa\Omega>1$) towards
an optomechanically-induced optical parametric oscillator. Already
below threshold, it modifies the optical band in a distinct way, as
can be seen in the highlighted region of Fig.~\ref{fig:Optomechanical-band-structure}(b)
(gray box), shown as a close-up in Figs.~\ref{fig:Optomechanical-band-structure}(c-d).
Above threshold, the array produces beams with opposite quasimomentum,
entangled in their quadratures. Their intensity increases smoothly
with the laser power while the intensity of the laser-driven pump
mode at $k=0$ saturates. This optical instability does not have any
analogue in single-mode optomechanics.

\begin{figure}
\includegraphics[width=1\columnwidth]{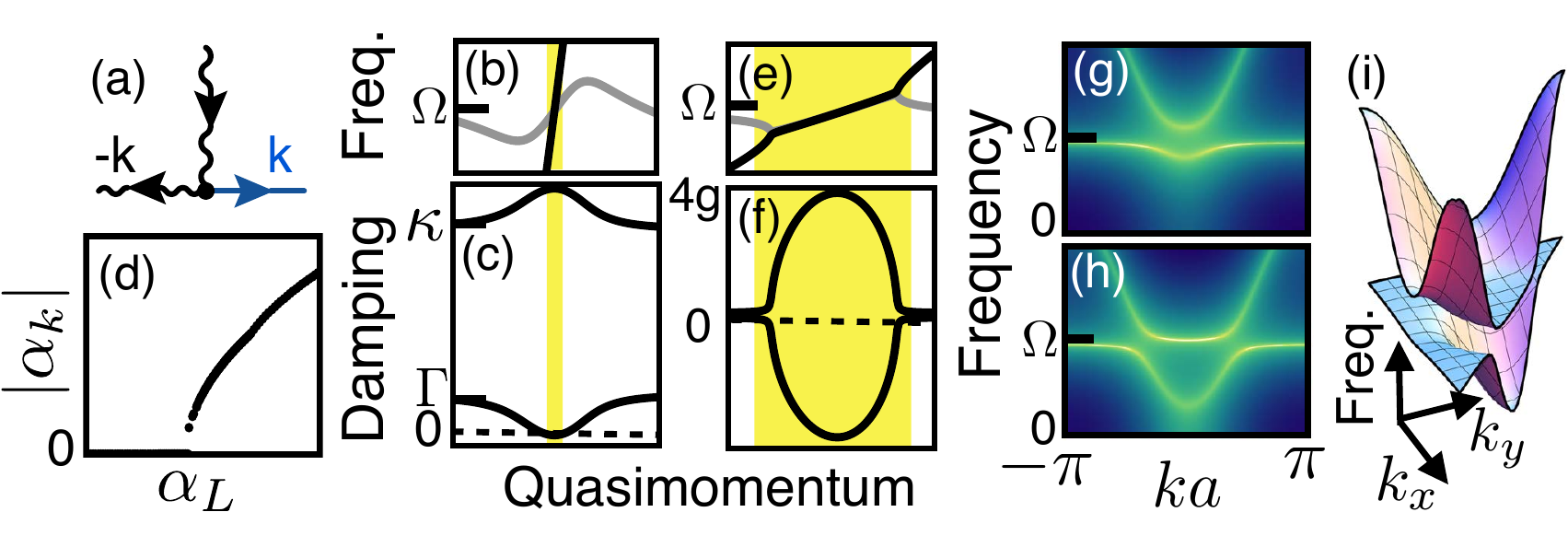}

\caption{\label{fig:bandstructure-bluedet}Band structure in the blue- and
red-detuned regime, for an optomechanical array: (a-f) Close to the
blue detuned sideband, $\Delta_{k}\approx\Omega$, a drive photon
is parametrically converted into a photon-phonon pair with opposite
quasimomenta; see Feynman diagram (a). Panels (b) and (c) show the
associated frequencies and damping rates in the weak coupling regime
$g\ll\kappa$, near the instability. The unstable region, where the
damping rates become negative, is marked in yellow. Slightly above
the instability threshold, a pair of optical and mechanical down-converted
beams are generated. The typical bifurcation behaviour of the optical
amplitude $|\alpha_{k}|$ is shown as a function of the driving amplitude
$|\alpha_{L}|$ in (d) {[}note $\left|\beta_{k}\right|$ looks similar{]}.
Panels (e, f) show the same for the strong coupling regime $g\gg\kappa$.
Here we find the array dynamics becomes chaotic, and the bandstructure
thus is to be understood to describe only the initial, transient dynamics
after a quench of parameters. (g,h,i): Close to the red detuned sideband,
$-\Delta_{k}=\Omega$, the optical and mechanical bands hybridize,
forming polariton excitations; see also \cite{Chang2011,Duan2012}.
When the unperturbed photon band touches the mechanical band at $k=0$,
the low momentum polaritons have effective mass $\hbar/Ja^{2}$, much
smaller than the phonon mass; see panel (g) ($\Delta=-2\Omega$, the
other parameters are the same as in Fig.~2(b)). For a 2D array, polaritons
may form on a closed contour in momentum space, see the cut panel
(h) and the 3D plot panel (i).}
\end{figure}

We note that an array version of optomechanical self-oscillations,
generated for blue-detuned drive, also exists (Fig.~\ref{fig:bandstructure-bluedet}a-f),
while strong-coupling physics on the red-detuned side leads to the
formation of optomechanical polaritons (Fig.~\ref{fig:bandstructure-bluedet}g-i).

\begin{figure}
\includegraphics[width=1\columnwidth]{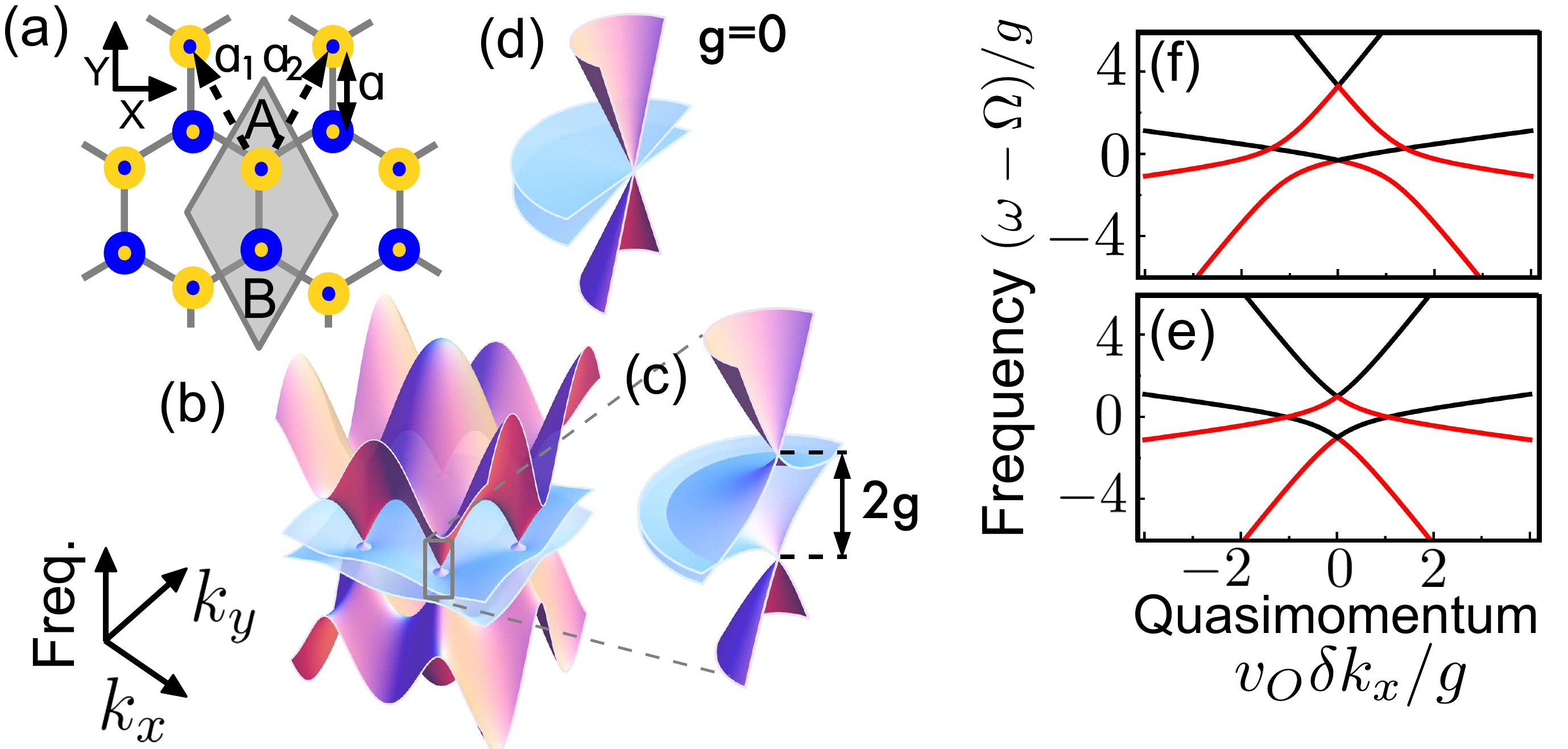}

\caption{Optomechanical Dirac Physics: (a) Sketch of the optomechanical honeycomb
lattice. The unit cell contains the optomechanichal cells A and B.
The corresponding sublattices are generated by discrete translations
by the lattice basis vectors $\vec{a}_{1}$ and $\vec{a}_{2}$. (b)
3D view of the optomechanical band structure at the red detuned sideband,
$\Omega=-\tilde{\Delta}$. (c-d) Close-up close to a symmetry point
for the interacting and unperturbed case, respectively. (e) Cut of
the band structure for the same parameters ($v_{M}=0.1v_{O}$) and
(f) for large detuning $|\tilde{\Delta}|-\Omega=3g$ where an avoided
crossing appears. The bands with opposite helicity (plotted in different
colors) display exact crossings.}
\label{fig:diraccones}
\end{figure}

Hitherto, we have described the behavior in simple lattices. We now
turn to an optomechanical honeycomb lattice, see the sketch in Fig.~\ref{fig:diraccones}(a).
The tight-binding model for this non-Bravais lattice is well-known
for graphene \cite{Neto2009}, but is now also studied for photonic
crystals \cite{Ochiai2009,Rechtsman2012}. The band structure includes
special points where the upper and lower bands touch, forming so-called
Dirac cones, which are robust, topologically protected structures.
There, both the optical and mechanical band are described by a relativistic
massless 2D Dirac equation, but with different velocities: fast photons
(with velocity $v_{O}=3aJ/2$) and slow phonons (velocity $v_{M}=3aK/2$).
For concreteness, we focus on excitations around the symmetry point
$\vec{K}=2\pi(3^{-1/2},1)/3a$ and consider the quasimomentum $\delta\vec{k}$
relative to that point. As usual, one can assign a binary degree of
freedom $\hat{\sigma}_{z}=\pm1$ to excitations on the A/B sublattices
and write the Hamiltonian (for either of the bands) as $\hbar v\hat{\vec{\sigma}}\cdot\delta\vec{k}=\hbar v(\hat{\sigma}_{x}\delta k_{x}+\hat{\sigma}_{y}\delta k_{y})$.
To investigate the effects of optomechanical interaction, we focus
on the red-detuned sideband $|\tilde{\Delta}+\Omega|\ll\Omega$, where
the optical and mechanical cones come close, in the strong coupling
regime $g\gg\kappa$, where dissipative effects are less important.
We introduce a second binary degree of freedom, $\hat{\tau}_{z}=\pm1$,
to denote optical/mechanical excitations. This could be viewed as
analogous to the electron spin, but contrary to the usual situation,
the velocity depends strongly on this spin degree of freedom. We find
the optomechanical Dirac Hamiltonian \cite{SMoptoarray} 
\begin{equation}
\hat{H}_{D}/\hbar=\bar{\omega}+\delta\omega\hat{\tau}_{z}/2-(\bar{v}+\delta v\hat{\tau}_{z}/2)\hat{\vec{\sigma}}\cdot\delta\vec{k}-g\hat{\tau}_{x}.
\end{equation}
where we have introduced the parameters $\bar{\omega}=(\Omega+|\tilde{\Delta}|)/2,$
$\delta\omega=|\tilde{\Delta}|-\Omega$, $\bar{v}=(v_{O}+v_{M})/2$,
and $\delta v=v_{O}-v_{M}$. (Note that $\hat{H}_{D}$ represents
a single-particle Hamiltonian for the photon/phonon excitations which
could be turned into second-quantized form for $\hat{a}_{k},\hat{b}_{k}$
in the usual manner.) The Dirac Hamiltonian is valid for $|\delta\vec{k}|\ll a^{-1}$.
In stark contrast to the standard case, the resulting optomechanical
Dirac band structure is dispersive 
\begin{equation}
\omega_{\tau,\sigma}=\bar{\omega}-\sigma\bar{v}|\delta\vec{k}|+\tau[g^{2}+(\delta\omega-\sigma\delta v|\delta\vec{k}|)^{2}/4]^{1/2}.
\end{equation}
Close to the symmetry point, it consists of two pairs of cones shifted
in frequency by $\sqrt{\delta\omega^{2}+g^{2}}.$ By varying $\delta\omega$
from a large positive value to a large negative one, the velocity
of the upper pair of cones smoothly goes from $v_{O}$ to $v_{M}$
as the corresponding excitations turn from optical to mechanical.
The frequency shift has its minimum value $g$ at the red detuned
sideband $\delta\omega=0$ where both cones describe polaritons with
velocity $\bar{v}$, see Fig.~\ref{fig:diraccones}(e). For large
detunings $\delta\omega\gg g$, the cones with same helicity $\hat{\vec{\sigma}}\cdot\delta\vec{k}/|\delta\vec{k}|$
hybridize away from the symmetry point displaying an avoided crossing
for $v_{O}|\delta\vec{k}|\sim\delta\omega$ (f). Notice that the optomechanical
interaction mixes the optical and the mechanical bands but, as in
the standard Dirac Hamiltonian, the helicity remains conserved. This
symmetry is responsible for the band crossings in Fig.~4.

In finite honeycomb arrays, photon/phonon edge states would appear,
depending on the type of edge, as in graphene. A potential landscape
can be introduced by modifying locally the properties of the optomechanical
cells, which would give rise to effects such as Klein tunneling.

\begin{figure}
\includegraphics[width=8cm]{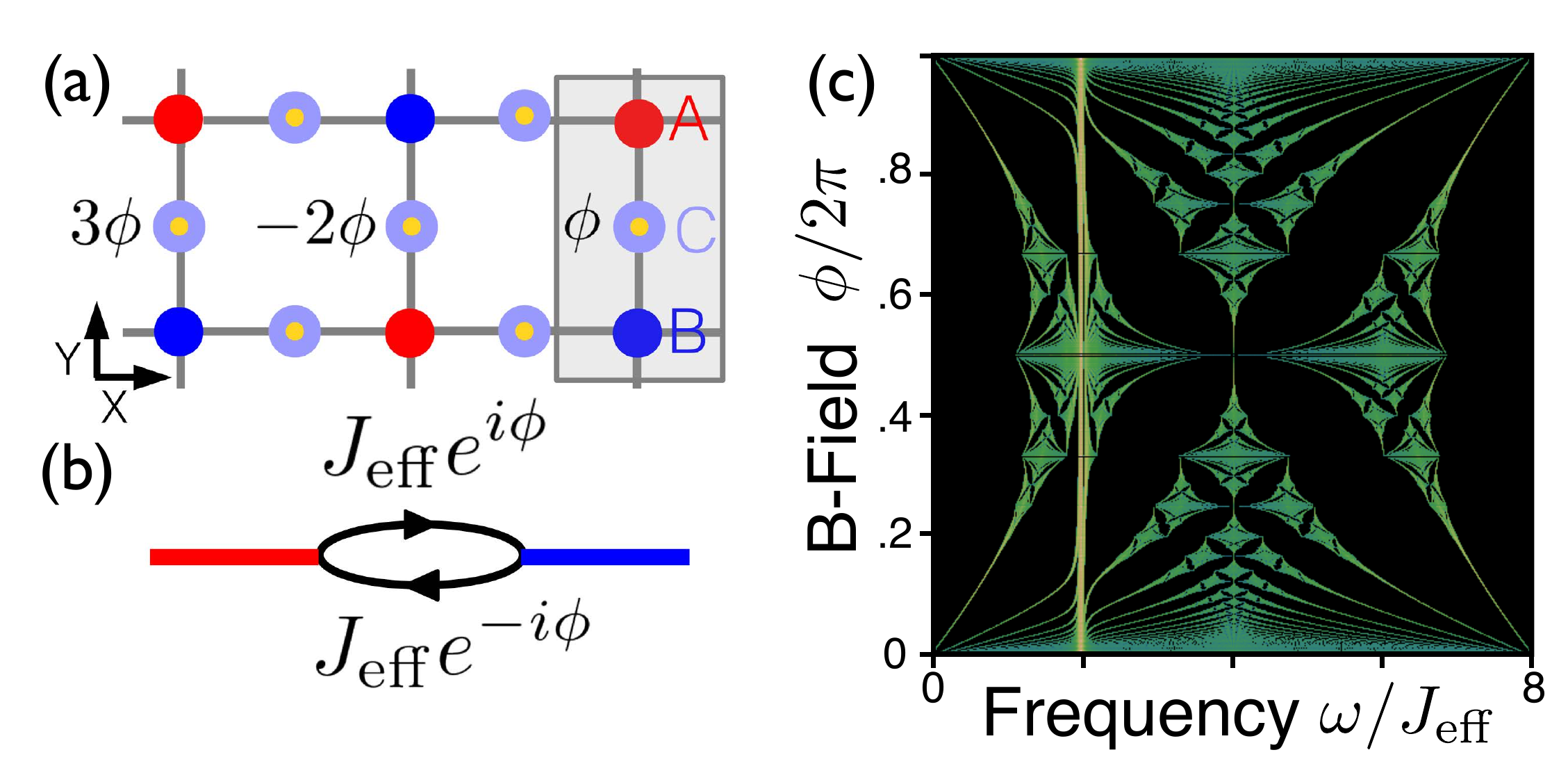}\caption{\label{fig:Gauge field}Optomechanical photonic Gauge fields: (a)
Sketch of the optomechanical realization of a gauge field for photons.
A photon hopping from A to B picks up a phase set by the phase of
the mechanical oscillations on the corresponding optomechanical link.
The phase configuration indicated in the figure corresponds to a constant
effective magnetic field $-\phi/a^{2}$. (b) Effective coupling of
the quasienergy levels involved in a hopping transition in the shaded
region. (c) Optomechanical density of states in presence of the effective
magnetic field and an additional red detuned onsite driving leading
to the hybridization of the Hofstadter butterfly and a mechanical
band (bright line) {[}Parameters: $\Delta_{A}=-2\Omega$ $K=0.005\Omega$,
$J_{{\rm eff}}=0.5\Omega$, $g=0.1\Omega${]}.}
\end{figure}
We now show that by adopting a slightly different lattice topology
and time-dependent laser driving, one can engineer an artificial magnetic
field for the photons inside an optomechanical array. Recently, it
has been predicted that an effective gauge field $A_{{\rm eff}}$
can be introduced in a photonic square lattice comprising two sublattices
$A$ and $B$ by electro-optically modulating the nearest-neighbor
coupling constants: $J_{ij}=J\cos(\omega_{{\rm ext}}t+\phi_{ij})$
\cite{Fang2012}. For resonant modulation, $\omega_{{\rm ext}}=\omega_{A}-\omega_{B}$,
a photon hopping from site $i$ on A to a neighboring site $j$ on
B picks up the phase of the modulation $\phi_{ij}$like a charged
particle subjected to the gauge potential $\int_{r_{i}}^{r_{j}}A_{{\rm eff}}\cdot dr=\phi_{ij}$.
In our proposal, the nearest neighbor coupling between the optical
sublattices $A$ and $B$ would be mediated by an intermediate lattice
$C$; see Fig.~\ref{fig:Gauge field}(a). Driving the optomechanical
cells on this lattice $C$ with a sinusoidal time-dependent power
gives rise to an oscillating radiation pressure force and thus classical
(large-amplitude) oscillations of the mechanical mode at the link:
$\beta(t)=\beta\exp(-i\omega_{{\rm ext}}t)$. These oscillations weakly
modulate the eigenfrequency $\omega_{C}(t)$ of an optical mode $\hat{a}_{C}$
at the link: $\omega_{C}(t)=\bar{\omega}_{C}+2g_{0}|\beta|\cos(\omega_{{\rm ext}}t+\phi)$,
$g_{0}|\beta|\ll\omega_{{\rm ext}}$. The effective coupling $J_{{\rm eff}}\exp[i\phi]$
between $\hat{a}_{A}$ and $\hat{a}_{B}$ is mediated by virtual tunneling
through the modulated mode $\hat{a}_{C}$ (with coupling constants
$J_{A}$ and $J_{B}$, respectively) accompanied by the exchange of
a phonon. Here, $\phi$ is set by the phase of the laser power modulation
whereas $J_{{\rm eff}}=g_{0}|\beta|J_{A}J_{B}/[2(\omega_{A}-\omega_{C})(\omega_{B}-\omega_{C})]$
\cite{SMoptoarray}. 

Creating the artificial magnetic field for the photons requires modulating
the incoming laser intensity temporally at each link cell, with different
modulation phases on different links. A naive approach for very few
cells would make use of laser beams whose intensity is individually
controlled (either with tightly focused beams or injected along wave
guides). However, the same effect can be obtained with no more than
two laser beams for an entire large optomechanical array, provided
that wave front engineering is applied. Suppose the 'carrier' beam
amplitude is $E_{1}=E_{10}e^{-i\omega_{L}t}$ and the 'modulation'
beam amplitude is $E_{2}=E_{20}e^{-i(\omega_{L}+\omega_{{\rm ext}})t-i\phi(x,y)}$,
where $\phi$ is an imprinted phase shift depending on the position
within the array plane. Then the intensity, and thus the local radiation
pressure force driving the mechanical oscillations, is $\left|E_{10}\right|^{2}+\left|E_{20}\right|^{2}+2{\rm Re}[E_{10}^{*}E_{20}e^{-i(\omega_{{\rm ext}}t+\phi(x,y))}]\,,$
which is what was needed. For the resulting mechanical oscillations
$\beta(t)$, there will be an extra, but constant and thus irrelevant
phase shift depending on the relation between $\omega_{{\rm ext}}$
and the mechanical frequency $\Omega$ on the links. For resonant
drive, the amplitude $\left|\beta\right|$ on the links will be enhanced
by the mechanical quality factor, and thus much larger than any spurious
amplitude on other cells (which can be chosen off-resonant with $\omega_{{\rm ext}}$),
which could be further suppressed by engineering the intensity pattern
$\left|E_{20}(x,y)\right|^{2}$ as well. 

 Our proposal demonstrates how optomechanical systems can contribute
to the recent efforts in creating gauge fields for photons \cite{Hafezi2011,Umucallar2011,Bermudez2012,Fang2012,Rechtsman2012}.
The resulting optomechanical Hofstadter butterfly fractal level scheme
is displayed in Fig.~\ref{fig:Gauge field}, for a square lattice.
The photonic magnetic field can be implemented in arbitrary lattices
(like the honeycomb lattice) and it could be combined with magnetic
fields for phonons, based on \cite{Bermudez2011,Habraken2012}.

 The realizability of optomechanical arrays based on optomechanical
crystals has been shown by theoretical studies including ab-initio
simulations \cite{Safavi-Naeini2010,2013_Ludwig}. Moreover, nanocavity
arrays in photonic crystals have been demonstrated already in experiments
\cite{2008_notomilargephotonarrays}. The parameters for the phenomena
analyzed here are within experimental reach and we estimate the effects
of disorder to be minor for realistic array sizes, see Supplemental
Material \cite{SMoptoarray}. 

In summary, we have shown how to tailor the flow of photons and phonons
in an optomechanical metamaterial, whose band structure can be tuned
via the driving laser.  Future studies could deal with the rich nonlinear
dynamics of competing unstable modes, the additional effects brought
about by nonlinearities on the single-photon level, the influence
of more complex spatial patterns imprinted by the driving laser, or
photon/phonon-transport phenomena.

\textbf{Acknowledgements}

This work was supported via an ERC Starting Grant OPTOMECH and via
the DARPA program ORCHID.

\emph{}

\emph{}

\clearpage

\global\long\def\theequation{S.\arabic{equation}}
 \global\long\def\thefigure{S\arabic{figure}}

\thispagestyle{empty}
\onecolumngrid
\begin{center}
{\fontsize{12}{12}\selectfont \textbf{Supplemental material for the article: ``Optomechanical Metamaterials:
Dirac polaritons, Gauge fields, and Instabilities''\\[5mm]}}
{\normalsize M Schmidt$^{1}$, V Peano$^{1}$ and F Marquardt$^{1,2}$\\[1mm]}
{\fontsize{9}{9}\selectfont
\textit{	$^1$University of Erlangen-N\"urnberg, Staudtstr. 7, Institute for Theoretical
Physics, D-91058 Erlangen, Germany\\
 			$^2$Max Planck Institute for the Science of Light, G\"unther-Scharowsky-Stra\ss e
1/Bau 24, D-91058 Erlangen, Germany}}
\vspace*{6mm}
\end{center}
\normalsize
\twocolumngrid

\section*{Stationary states and Hamiltonian in momentum space}

The steady state light amplitude $\alpha=\langle\hat{a}_{j}\rangle$
and mechanical displacements $\beta=\langle\hat{b}_{j}\rangle$ can
be immediately computed by plugging the mean field ansatz in the full
nonlinear equations of motion. The resulting equations have the same
form as the standard equations for a single optomechanical cell \cite{SM_Meystre1985}
\begin{equation}
\alpha=\alpha_{L}/[\Delta+2g_{0}\beta-\nu_{O}+i\kappa/2],\qquad\beta=g_{0}|\alpha|^{2}/(\Omega+\nu_{M}).\label{eq:stationaryoscillations}
\end{equation}
As it is well known from the standard case of single-mode optomechanics,
the eigenfrequency of the optical mode is shifted by the radiation
pressure (in the main text we have incorporated this shift in the
effective detuning $\tilde{\Delta}=\Delta+2g_{0}\beta$). Moreover,
the optical and mechanical eigenfrequencies are shifted by the coupling
to the modes in the neighboring cells. The dependence of the corresponding
eigenfrequency shifts $\nu_{O}$ and $\nu_{M}$ on the hopping term
in the Hamiltonian is discussed below. In the main text, we have considered
a real positive amplitude $\alpha$. This does not imply any loss
of generality as the phase of $\alpha$ can be fixed by a global gauge
transformation when all the cells are driven with the same phase. 

For $|\alpha_{L}|>[\kappa^{3}(\Omega+\nu_{M})/(6\sqrt{3}g_{0}^{2})]^{1/2}$,
there exists a $\Delta$-interval where Eq.~(\ref{eq:stationaryoscillations})
has three solutions. As it is well known from the standard case of
single-mode optomechanics, one of the three solutions is always unstable.
On the other hand, the stability analysis of the remaining two solutions
is specific of the optomechanical metamaterial and is carried out
in the main text. Since for appropriate values of the damping rates
there are two stable solutions, it is possible to observe a hysteresis
of the band structure as a function of the driving parameters ($\Delta$
or $\alpha_{L}$).

In a Bravais lattice, the stationary solutions depend on the lattice
geometry only via the eigenfrequency shifts $\nu_{O}$ and $\nu_{M}$.
For the nearest neighbor hopping considered in the main text we have
$\nu_{O}=-zJ$ and $\nu_{M}=-zK$, where $z$ is the coordination
number. On the other hand, for a general short range hopping
\[
\hat{H}_{\mathbf{{\rm hop}}}=-\hbar\sum_{ij}J(i-j)\hat{a}_{j}^{\dagger}\hat{a}_{i}+K(i-j)\hat{b}_{j}^{\dagger}\hat{b}_{i}+c.c.,
\]
we find $\nu_{O}=-\sum_{r}J(r)$ and $\nu_{M}=-\sum_{r}K(r)$. 

The Hamiltonian in momentum space Eq.~(2) of the main text follows
directly from the plane-wave ansatz for the Bloch modes and the linearization
of the optomechanical interaction
\[
\hat{a}_{j}^{\dagger}\hat{a}_{j}(\hat{b}_{j}+\hat{b}_{j}^{\dagger})\approx{\rm const}+2\beta\delta\hat{a}_{j}^{\dagger}\delta\hat{a}_{j}+\alpha(\delta\hat{a}_{j}+\delta\hat{a}_{j}^{\dagger})(\delta\hat{b}_{j}+\delta\hat{b}_{j}^{\dagger})
\]
where $\delta\hat{a}_{j}=\hat{a}_{j}-\alpha$ and $\delta\hat{b}_{j}=\hat{b}_{j}-\beta$.

\section*{Optomechanical band structure of an array on a honeycomb lattice}

The two sublattices of a honeycomb lattice are mapped onto each other
by a rotation by a $\pi$-angle around the center of the unit cell,
a transformation belonging to the crystallographic point group, see
Fig.~4(a) of the main text. Due to this symmetry, for a driving below
a finite threshold, the stationary light amplitudes and mechanical
displacements are independent not only of the cell but also of the
sublattice, $\alpha=\langle\hat{a}_{is}\rangle$ and $\beta=\langle\hat{b}_{is}\rangle$
(index $s$ indicates the sublattice $s=1,2$ correspond to sublattice
A and B, respectively). In the tight-binding approximation, they are
given by Eq.~(\ref{eq:stationaryoscillations}) with $\nu_{O}=-3J$
and $\nu_{M}=-3K$. 

Since we describe both photons and phonons in the tight-binding approximation,
their Bloch modes have the same wavefunctions $|\psi_{s}(\vec{k})\rangle e^{i\vec{k}\cdot r_{j}}$
(the eigenfunctions of the tight-binding model of electrons in graphene
\cite{SM_PhysRev.71.622}). They are defined by 
\begin{eqnarray*}
\delta\hat{a}_{js} & = & N^{-1/2}\sum_{\vec{k}s'}e^{i\vec{k}\cdot \vec{r}_{j}}\langle s|\psi_{s'}(\vec{k)}\rangle\hat{a}_{\vec{k}s'},\\
\hat{H}_{{\rm hop}} & = & -\hbar\sum_{\vec{k}s}f_{s}(\vec{k})\left(J\hat{a}_{\vec{k}s}^{\dagger}\hat{a}_{\vec{k}s}+K\hat{b}_{\vec{k}s}^{\dagger}\hat{b}_{\vec{k}s}\right)+{\rm const}.
\end{eqnarray*}
and likewise for $\hat{b}_{\vec{k}s}$ ($|s=1,2\rangle$ indicate
an excitation on sublattice A and B, respectively). From the above
definitions and the definition of $\hat{H}_{{\rm hop}}$ (in the main
text), one immediately derives the eigenvalue problem
\[
f_{s}(\vec{k})|\psi_{s}(k)\rangle=[d'(\vec{k})\hat{\sigma}_{x}-d''(\vec{k})\hat{\sigma}_{y}]|\psi_{s}(\vec{k})\rangle,
\]
where $d'(\vec{k})$ and $d''(\vec{k})$ are the real and imaginary
part of $d(\vec{k})=1+e^{i\vec{k}\cdot \vec{a}_{1}}+e^{i\vec{k}\cdot \vec{a}_{2}}$
and $\hat{\vec{\sigma}}$ is a set of Pauli matrices, $\hat{\sigma}_{z}|s\rangle=(-)^{s+1}|s\rangle$.
Notice that the linearized optomechanical interaction is then diagonal
on the common Bloch basis of photons and phonons 
\[
\sum_{js}(\delta\hat{a}_{js}+\delta\hat{a}_{js}^{\dagger})(\delta\hat{b}_{js}+\delta\hat{b}_{js}^{\dagger})=\sum_{\vec{k}s}(\hat{a}_{\vec{k}s}+\hat{a}_{-\vec{k}s}^{\dagger})(\hat{b}_{-\vec{k}s}+\hat{b}_{\vec{k}s}^{\dagger}).
\]
Hence, equation (4) of the main text for the optomechanical band structure
of a Bravais lattice applies also to each band of the honeycomb lattice
separately. In this case, $\Delta_{\vec{k}s}=\tilde{\Delta}+Jf_{s}(\vec{k})$
and $\Omega_{\vec{k}s}=\Omega-Kf_{s}(\vec{k})$ with $\tilde{\Delta}=\Delta+2g^{2}/(\Omega-3K)$
and $f_{s}=(-)^{s}|d(\vec{k})|$. 

In the strong coupling limit $g\gg\kappa$ where dissipative effects
are less important and for red detuned driving $-\Delta_{\vec{k}}\gg g$
$ $where the two-mode squeezing part of the interaction is negligible,
the system is described by a particle-conserving second-quantized
Hamiltonian. In this regime, we can switch to a single-particle picture
introducing the polariton eigenfunctions $|\psi_{\ell}(\vec{k})\rangle e^{i\vec{k}\cdot r_{j}}$
and eigenenergies $\hbar\omega_{\ell}(\vec{k})$. They are defined
by
\begin{eqnarray*}
\delta\hat{a}_{js}(t) & \approx & N^{-1/2}\sum_{\vec{k}\ell=1,\dots,4}e^{i[\vec{k}\cdot \vec{r}_{j}-\omega_{\ell}(\vec{k})t]}\langle1,s|\psi_{\ell}(\vec{k})\rangle\hat{c}_{\vec{k}\ell}\\
\delta\hat{b}_{js}(t) & \approx & N^{-1/2}\sum_{\vec{k}\ell=1,\dots,4}e^{i[\vec{k}\cdot \vec{r}_{j}-\omega_{\ell}(\vec{k})t]}\langle2,s|\psi_{\ell}(\vec{k})\rangle\hat{c}_{\vec{k}\ell}.
\end{eqnarray*}
 On the left-hand side, $\delta\hat{a}_{js}(t)$ and $\delta\hat{b}_{js}(t)$
are solutions of the corresponding Heisenberg equations (the approximation
consists in neglecting the two-mode squeezing part of the optomechanical
interaction). On the right-hand side, the quantum numbers in $|s',s\rangle$
$ $denotes optical/mechanical excitations ($s'=1/2$) on sublattice
A/B ($s=1/2$). From the Heisenberg equations for $\delta\hat{a}_{js}(t)$
and $\delta\hat{b}_{js}(t)$, we derive the time-independent single
particle Schr\"odinger equation $\hbar\omega_{\ell}(k)|\psi_{\ell}\rangle=\hat{H}_{H}|\psi_{\ell}\rangle$
where
\[
\hat{H}_{H}/\hbar=\bar{\omega}+\delta\omega\hat{\tau}_{z}/2-(\bar{v}+\delta v\hat{\tau}_{z}/2)[d'(\vec{k}){\rm \hat{\sigma}_{x}}+d''(\vec{k})\hat{\sigma}_{y}]-g\hat{\tau}_{x}.
\]
Here, we have introduced the parameters $\bar{\omega}=(\Omega+|\tilde{\Delta}|)/2,$
$\delta\omega=|\tilde{\Delta}|-\Omega$, $\bar{v}=(v_{O}+v_{M})/2$,
and $\delta v=v_{O}-v_{M}$, and the set of Pauli matrices $\hat{\vec{\tau}}$,
$\hat{\tau_{z}}|s',s\rangle=(-)^{1+s'}|s',s\rangle$. The spectrum $\omega_{\ell}$
is shown in Fig.~4 of the main text. By expanding $d(\vec{k})$ around
the symmetry point $\vec{K}=2\pi(3^{-1/2},1)/3a$ we arrive to the
optomechanical Dirac equation $\hat{H}_{D}$.

\section*{Generalization to multiband lattices}

The above analysis applies also to any $symmetric$ multiband lattice,
that is a lattice formed by sublattices any of which can be mapped
to another by a transformation belonging to the crystallographic point
group, e.~g.~ an optomechanical array on a Kagome lattice (which
has three bands). Due to the symmetry, the stationary light amplitudes
and displacements are independent of the sublattice and the cell,
$\alpha=\langle\hat{a}_{is}\rangle$ and $\beta=\langle\hat{b}_{is}\rangle$
($s=1,\dots,M$). They are given by Eq.~(\ref{eq:stationaryoscillations})
with $\nu_{O}=-\sum_{rs'}J_{ss'}(r)$ and $\nu_{M}=-\sum_{rs'}K_{ss'}(r)$.
For nearest neighbor hopping, photons and phonons have the same Bloch
wavefunctions and the band index $s$ is a conserved quantity. Therefore,
formula (4) of the main tex for the band structure applies to each
band separately. The coupling to non-nearest neighbor sites can be
regarded as a perturbation introducing a small interaction between
bands with different band index $s$ which leads to additional resonances
where the optical-vibrational mixing is enhanced. For a general multiband
lattice, the stationary solutions depend on the sublattice and the
coupling between different bands is not necessarily small.

\section*{Derivation of the Langevin equations}

The Langevin equations (3) of the main text follow by assuming the
weak linear coupling to a Markovian heat bath \cite{SM_Walls2008}. The
latter describes other mechanical and electronic degrees of freedom
in the sample (for the phononic Bloch modes $\hat{b}_{k}$) and the
electromagnetic field in free space (for the photonic Bloch modes
$\hat{a}_{k}$). If we assume an infinitely extended system with discrete
translational symmetry in the xy-plane, then the in-plane symmetry
under discrete translation by the lattice unit vectors is preserved
in the full Hamiltonian $\hat{H}$ which incorporate also these degrees
of freedom. Therefore we will assume that Bloch modes with different
quasimomentum are coupled to independent baths. With this assumption,
we find the noise correlators $\langle\hat{\xi}_{k}(t)\hat{\xi}_{k'}^{\dagger}(0)\rangle=\kappa\delta_{k,k'}\delta(t)$,
$\langle\hat{\eta}_{k}(t)\hat{\eta}_{k'}^{\dagger}(0)\rangle=\Gamma(\bar{n}+1)\delta_{k,k'}\delta(t)$
and $\langle\hat{\eta}_{k}^{\dagger}(t)\hat{\eta}_{k'}(0)\rangle=\Gamma\bar{n}\delta_{k,k'}\delta(t)$.
Here $\bar{n}$ is the bosonic occupation number, $\bar{n}=(\exp[\hbar\Omega/k_{B}T]-1)^{-1}$.
In the generic case, the relaxation rates $\Gamma$ and $\kappa$
depend on the quasimomentum. Quasimomentum independent relaxation
rates correspond to independent fluctuations on different lattice
sites.

Notice that the Langevin equations for a general multi-band array
are symmetric under the substitution $k\to-k$. As a consequence,
it is useful to consider the Bloch mode amplitudes $\hat{X}_{ks}^{(a)}=(\hat{a}_{ks}+\hat{a}_{-ks}^{\dagger})/\sqrt{2}$
and $\hat{X}_{ks}^{(b)}=(\hat{b}_{ks}+\hat{b}_{-ks}^{\dagger})/\sqrt{2}$.
These amplitudes are actually the plane-wave coefficients of the fields
$(\delta\hat{a}_{j}+\delta\hat{a}_{j}^{\dagger})/\sqrt{2}$ and $(\delta\hat{b}_{j}+\delta\hat{b}_{j}^{\dagger})/\sqrt{2}$.
Here, we are interested ultimately in obtaining the eigenfrequencies
defining the band structure. These can be obtained from the homogeneous
part of the Langevin equations, i.e. the equations that result when
taking the average and eliminating the noise terms. Thus, we now consider
only the expectation values, $X_{ks}^{(a)}=\left\langle \hat{X}_{ks}^{(a)}\right\rangle $,
likewise for $X_{ks}^{(b)}$. Since the coefficients in the resulting
set of coupled linear equations are real-valued, a monochromatic solution
with complex eigenfrequency $\omega_{k}$ must be accompanied by the
complex conjugated solution with eigenfrequency $-\omega_{k}^{*}$
(monochromatic solutions describing overdamped excitations can have
purely imaginary eigenfrequencies). In the special case of a Bravais
lattice (or a symmetric multi-band lattice with nearest neighbor hopping)
the equations of motion for these quadratures are
\begin{eqnarray}
\ddot{X}_{k}^{(a)} & = & -(\Delta_{k}^{2}+\kappa^{2}/4)X_{k}^{(a)}-\kappa\dot{X}_{k}^{(a)}-2\Delta_{k}gX_{k}^{(b)},\nonumber \\
\ddot{X}_{k}^{(b)} & = & -(\Omega_{k}^{2}+\Gamma^{2}/4)X_{k}^{(b)}-\Gamma\dot{X}_{k}^{(b)}+2\Omega_{k}gX_{k}^{(a)}.
\end{eqnarray}
In this case, the eigenfrequency spectrum consists of the solutions
of Eq.~(4) of the main text. A graphical study of the polynomial
equation shows that two scenarios are possible: i) either there are
two pairs of opposite complex conjugate eigenfrequencies or ii) one
pair and two purely imaginary eigenfrequencies.

\section*{Photon emission spectrum}

From input output formalism, the intensity of the radiation emitted
at frequency $\omega_{L}-\omega$ and with inplane quasimomentum $k$
is proportional to the noise spectrum $S(k,\omega)\equiv\int dt\exp[i\omega t]\langle\hat{a}_{k}^{\dagger}(t)\hat{a}_{k}\rangle$.
By plugging the solution of the Langevin equations into the definition
of $S(k,\omega)$ and evaluating the correlators of the noise forces,
we find
\begin{equation}
S(k,\omega)=\frac{4\kappa g^{4}\Omega^{2}+\Gamma\sigma_{{\rm M}}(\omega)}{|{\cal N}(\omega)|^{2}}\label{eq:powerspectrum}
\end{equation}
in terms of the analytical functions
\begin{eqnarray*}
 &  & \sigma_{{\rm M}}=g^{2}|\chi_{_{O}}(\omega)|^{-2}\left[(\bar{n}+1)|\chi_{_{M}}(-\omega)|^{-2}+\bar{n}|\chi_{_{M}}(\omega)|^{-2}\right]\\
 &  & {\cal N}(\omega)=[\chi_{_{O}}(\omega)\chi_{_{M}}(\omega)\chi_{_{O}}^{*}(-\omega)\chi_{_{M}}^{*}(-\omega)]^{-1}+4g^{2}\Delta_{k}\Omega_{k}.
\end{eqnarray*}
Here, we have introduced the free susceptibilities $\chi_{_{O}}(\omega)=[\kappa/2-i(\omega+\Delta_{k})]^{-1}$
and $\chi_{_{M}}(\omega)=[\Gamma/2-i(\omega-\Omega_{k})]^{-1}$. Notice
that ${\cal N}(\omega)$ coincides with the polynomial, in Eq.~(4)
of the main text, defining the band structure. When we extend $S(k,\omega)$,
which is defined for real-valued $\omega$, to the complex $\omega$-plane,
each eigenfrequency $\omega_{k}$ belonging to the band structure
is a second order pole of the photon emission spectrum. In the generic
situation, the spectrum consists of four Lorentzian peaks whose position
and FWHM correspond to the real and twice the imaginary part of the
corresponding eigenfrequency $\omega_{k}$.

Next, we derive a mapping between the noise correlators of the array
and the noise correlators of linearized single-mode optomechanics
(i.e. the standard system considered in the literature). We preliminary
observe that the Langevin equations (3) of the main text have the
same form as the standard Langevin equation for single mode optomechanics
\begin{eqnarray}
\dot{\hat{a}} & = & (i\Delta-\kappa/2)\hat{a}+ig(\hat{b}+\hat{b}^{\dagger})+\hat{\xi}\nonumber \\
\dot{\hat{b}} & = & (-i\Omega-\Gamma/2)\hat{b}+ig(\hat{a}^{\dagger}+\hat{a})+\hat{\eta}\,.\label{eq:linearizedeqofmotion-2_SM}
\end{eqnarray}
 The key difference is that in the equations for the array {[}(3)
of the main text{]} the pair of optical (mechanical) modes $\hat{a}_{k}$
and $\hat{a}_{-k}^{\dagger}$ ($\hat{b}_{k}$ and $\hat{b}_{-k}^{\dagger}$)
that are coupled by the optomechanical interaction are not a hermitian
conjugate pair (as the corresponding noise forces $\hat{\xi}_{k}$
and $\hat{\xi}_{k}^{\dagger}$ are not a hermitian conjugate pair).
This difference is not relevant while computing the dynamical form
factors as detailed below. Due to momentum conservation, the only
non-zero noise correlators are
\[
S_{cd}(k,\omega)\equiv\int dt\, e^{i\omega t}\langle\hat{c}_{k}(t)\hat{d}_{-k}\rangle
\]
 where $\hat{c}_{k},\hat{d}_{k}$ refer to either of $\hat{a}_{k},\hat{b}_{k},\hat{a}_{-k}^{\dagger},\hat{b}_{-k}^{\dagger}$.
Since the array modes $\hat{c}_{k},\hat{d}_{-k}$ are governed by
the same equations as the corresponding ladder operators $\hat{c}$
and $\hat{d}$ of single-mode optomechanics, $\hat{c}_{k}(t)$ ($\hat{d}_{-k}$)
is the same function of $\Delta_{k}$, $\Omega_{k}$, $\hat{\xi}_{k},$
$\hat{\eta}_{k}$, $\hat{\xi}_{-k}^{\dagger}$ and $\hat{\eta}_{-k}^{\dagger}$
($\Delta_{k}$, $\Omega_{k}$ $\hat{\xi}_{-k},$ $\hat{\eta}_{-k},$
$\hat{\xi}_{k}^{\dagger}$ and $\hat{\eta}_{k}^{\dagger}$ ) as $\hat{c}$
($\hat{d}$) of $\Delta$, $\Omega$, $\hat{\xi},$ $\hat{\eta,}$
$\hat{\xi}^{\dagger}$ and $\hat{\eta}^{\dagger}$. For equal phonon
bath temperature and decay rates $\kappa$ and $\Gamma$, also the
correlators of the relevant noise forces coincide. From this, we immediately
obtain the mapping between the noise spectra: 
\[
S_{cd}(k,\omega)=\left.\int dt\, e^{i\omega t}\langle\hat{c}(t)\hat{d}\rangle\right|_{\Delta=\Delta_{k},\Omega=\Omega_{k}}.
\]
where the correlator on the right-hand-side is the one calculated
for a single-mode optomechanical system. As Kubo formula expresses
the susceptibilities as the difference between two noise correlators,
a similar mapping obviously applies also to these functions. Hence,
an optomechanical metamaterial will display many phenomena known from
single-mode optomechanics, e.~g.~ sideband cooling, photon-phonon
entanglement, optomechanically induced transparency, etc.~\cite{SM_Aspelmeyer2013RMPArxiv}.
We emphasize though that in optomechanical arrays, phenomena that
usually occur for distinct parameter sets can coexist, since $\Delta_{k}$
varies in a finite range as a function of the quasimomentum. Note
also that the mapping of course does not extend to the full nonlinear
dynamics, where a wider range of momenta become mixed.

\section*{}

\section*{Details of the derivation of the optomechanical gauge field for photons }

We first consider the dynamics of the subsystem shown in the grey
shaded region in Fig.~5 (a). It comprises a cell on each sublattice,
A, B, C. This is sufficient to ultimately derive the effective coupling
between $A$ and $B$ to leading order in perturbation theory. We
denote the corresponding ladder operators by $\hat{a}_{A},$ $\hat{a}_{B},$
and $\hat{a}_{C}$, respectively. Since the second-quantized Hamiltonian
is particle conserving, it is most convenient to switch to the single-particle
picture. This means we solve the Schr\"odinger equation for the single-particle
wave function $\psi_{i}(t)\equiv(\alpha_{Ai}(t),\alpha_{Bi}(t),\alpha_{Ci}(t))$
of an excitation that can hop between the sites $A,B,C$ under the
influence of the driving field. Then, the equation of motion for $\psi$
reads $i\dot{\psi}=\hat{H}_{M}\psi$ with 
\[
\hat{H}_{M}=\begin{pmatrix}\omega_{A} & 0 & -J_{A}\\
0 & \omega_{B} & -J_{B}\\
-J_{A} & -J_{B} & \omega_{C}+2g_{0}|\beta|\cos(\omega_{{\rm ex}}t+\phi)
\end{pmatrix}
\]
The time periodicity ensures that there is a complete set of quasi-periodic
solutions of the Schr\"odinger equation $\psi_{i}(t+2\pi/\omega_{{\rm ex}})=\exp[-i2\pi\omega_{i}/\omega_{{\rm ex}}]\psi_{i}(t)$,
$i=1,2,3$. The quasienergies $\omega_{i,m}\equiv\omega_{i}+m\Omega$
and the time periodic functions $\phi_{i,m}\equiv\exp[i(\omega_{i}+m\Omega)t]\psi_{i}(t)$
{[}$m\in\mathbb{Z}${]} are eigenvalues and eigenvectors of the Floquet
Hamiltonian ${\cal H}\equiv-i\partial_{t}+\hat{H}_{M}$ \cite{SM_SAMBE1973}.

\begin{figure}

\includegraphics[width=1\columnwidth]{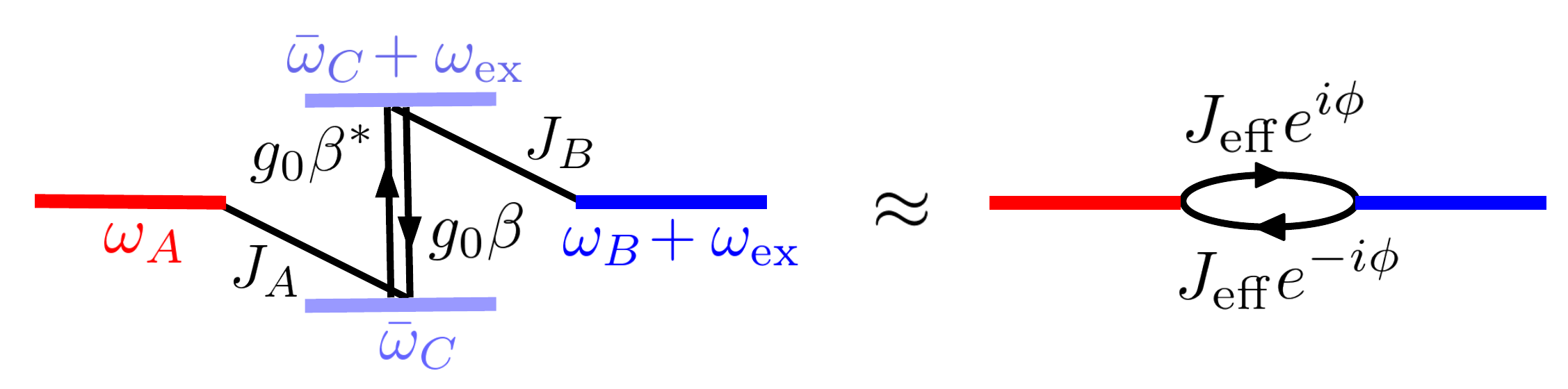}\caption{\label{fig:Floquet-level-scheme}Floquet level scheme describing the
hopping transitions between sites $A$ and $B$ mediated by the virtual
transition trough site $C$ accompanied by the exchange of a phonon.}

\end{figure}

We want to compute an effective Schr\"odinger equation describing the
resonant coupling of oscillations with frequency $\omega_{i}\sim\omega_{A}$
on site A and $\omega_{i}+\omega_{{\rm ex}}\sim\omega_{B}$ on site
B. Formally, we derive the effective Floquet Hamiltonian for the quasi-degenerate
quasienergy levels $\omega_{A}$ and $\omega_{B}+\omega_{{\rm ex}}$
(corresponding to the Floquet states $\hat{\phi}_{1,0}^{(0)}=(1,0,0)$
and $\hat{\phi}_{2,1}^{(0)}=(0,1,0)\exp[i\omega_{{\rm ex}}t]$) which
incorporates their coupling to leading order in $J_{A}/|\omega_{A}-\omega_{C}|$,
$J_{B}/|\omega_{B}-\omega_{C}|$, $g/\omega_{{\rm ex}}$. The leading
order process consists in a virtual hopping transition to site $C$,
the virtual absorption (emission) of a phonon and a virtual hopping
transition from site $C$, see sketch of the Floquet level scheme
in Fig.~\ref{fig:Floquet-level-scheme}.
 We consider the block of the Floquet Hamiltonian which includes only
the unperturbed quasienergy levels involved in such a process 
\[
{\cal \hat{H}}=\begin{pmatrix}\omega_{A} & 0 & -J_{A} & 0\\
0 & \omega_{B}+\omega_{{\rm ex}} & 0 & -J_{B}\\
-J_{A} & 0 & \omega_{C} & g_{0}\beta\\
0 & -J_{B} & g_{0}\beta^{*} & \omega_{C}+\omega_{{\rm ex}}
\end{pmatrix}.
\]
 Applying third order perturbation theory for quasi-degenerate levels,
we arrive at an effective block diagonal Floquet Hamiltonian. The
block which describes the dynamics of the light on sites $A$ and
$B$ reads 
\[
{\cal \hat{H}}_{{\rm eff}}=\begin{pmatrix}\tilde{\omega}_{A} & J_{{\rm eff}}e^{-i\phi}\\
J_{{\rm eff}}e^{i\phi} & \tilde{\omega}_{B}+\omega_{{\rm ex}}
\end{pmatrix}
\]
with $\tilde{\omega}_{A}=\omega_{A}+J_{A}^{2}/(\omega_{A}-\omega_{C})$,
$\tilde{\omega}_{B}=\omega_{B}+J_{B}^{2}/(\omega_{B}-\omega_{C})$
and $J_{{\rm eff}}=g_{0}|\beta|J_{A}J_{B}/[(\omega_{A}-\omega_{C})(\omega_{B}-\omega_{C})]$
. The Floquet Hamiltonian ${\cal H}_{{\rm eff}}$ is equivalent to
the time-dependent Hamiltonian
\[
\hat{H}_{{\rm eff}}=\begin{pmatrix}\tilde{\omega}_{A} & J_{{\rm eff}}e^{-i(\omega_{ex}t+\phi)}\\
J_{{\rm eff}}e^{i(\omega_{ex}t+\phi)} & \tilde{\omega}_{B}
\end{pmatrix}
\]
or in a frame rotating with frequency $\tilde{\omega}_{A}$ on site
$A$ and $\tilde{\omega}_{B}$ on site $B$ to the second-quantized
Hamiltonian $J_{{\rm eff}}e^{-i\phi}\hat{a}{}_{A}^{\dagger}\hat{a}{}_{B}+h.c.$
(for resonant driving, $\tilde{\omega}_{A}=\tilde{\omega}_{B}+\omega_{{\rm ex}}$
$ $). The same procedure can be applied to all nearest neighbors
sites leading to an effective hopping term between each nearest neighbors.
Hence, by appropriately tuning the driving frequency $\omega_{{\rm ex}}$
and phases $\phi_{ij}$, we can create the desired gauge field for
photons. 

In Fig.~5(c) we show how the standard Hofstadter Butterfly is modified
in presence of a red-detuned driving on each site (in addition to
the driving on the links which creates the effective magnetic field).
The resulting steady radiation enhances the optomechanical coupling
of each localized optical mode to a mechanical mode on the same cell
(with eigenfrequency $\Omega$). A equal detuning $\Delta$ of the
driving frequencies $\omega_{L}^{(A)}$ and $\omega_{L}^{(B)}$ on
sublattice $A$ and $B$, $\Delta=\omega_{L}^{(A)}-\tilde{\omega}_{A}=\omega_{L}^{(B)}-\tilde{\omega}_{B}$$ $
creates a monochromatic steady radiation on each cell $\alpha_{i}=\langle\hat{a}_{i}\rangle$
(in a frame rotating with the driving and within the RWA). Then, the
onsite fluctuations $\delta\hat{a}_{i}=\hat{a}_{i}-\alpha_{i}$ hops
in the effective magnetic field and are coupled to the corresponding
phononic modes by the standard linearized optomechanical interaction
with coupling constant $g_{i}\propto|\alpha_{i}|$. In the example
in Fig.~5 we have chosen $ $$g_{i}=g$. This is a good approximation
when the driving amplitude $\alpha_{L}$ is the same on all sites
and the detuning is larger than the effective coupling $J_{{\rm eff}}$
(then $\alpha_{i}\approx\alpha_{L}/(\Delta+i\kappa/2)$ with a small
magnetic-field induced site-dependent correction of order $\sim\alpha_{L}J_{{\rm eff}}/\Delta$).
Since the driving is red detuned we consider only the beam splitting
terms and the spectrum in Fig.~5(c) is computed by diagonalizing
the resulting single-particle optomechanical Hamiltonian. In particular,
we plot a smeared density of states averaged over a small area $d\phi\times d\omega$.

\section*{Discussion of the experimental realizability}

Single-mode optomechanical systems based on a single vibrational and
optical defect mode inside an optomechanical crystal have been experimentally
realized \cite{SM_Safavi-Naeini2010,SM_Gavartin2011PRL_OMC,SM_Chan2011Cooling}.
The reported experimental parameters are very promising, for example
$g_{0}=2.2\cdot10^{-4}\Omega$ \cite{SM_Chan2012} corresponds to a linearized
coupling strength $g\approx0.01\Omega$ for $2000$ photons circulating
in the cavity (a number reported in experiments \cite{SM_Chan2011Cooling}).
The side band resolved regime is routinely reached, with ratios $\kappa/\Omega$
in the range $0.01-0.1$ \cite{SM_Chan2011Cooling,SM_Chan2012} and mechanical
quality factors up to almost $Q\sim10^{6}$ were measured \cite{SM_Chan2012}. 

A number of theoretical investigations were carried out that indicate
the feasibility of 2D optomechanical crystals \cite{SM_Safavi-Naeini2010}
that support many optomechanically coupled modes with optical and
mechanical tunnel coupling \cite{SM_Heinrich2011CollDyn,SM_2013_Ludwig}
between neighboring sites as envisioned in this work. For defects
at a distance of a few lattice constants of the underlying photonic
crystal, one finds that the photon and phonon hopping amplitudes (J
and K) can reasonably reach values up to 10\% of the optical or vibrational
mode frequencies, respectively. This leads to typical ratios $J/K$
of about $10^{4}$ corresponding to a much larger speed of photon
propagation, unless special precautions are taken regarding the lattice
design. 

The observation of photon-phonon polaritons requires $4g>\kappa$
the so called strong coupling regime, which has been investigated
experimentally \cite{SM_Groblacher2009}. The regime of self-induced
mechanical oscillations requires $4g^{2}/\kappa\Gamma>1$, which is
also a prerequisite condition for side-band cooling and has been reached
experimentally with high quality vibrational modes and cryogenic cooling
\cite{SM_Teufel2011,SM_Chan2011Cooling}. The observation of the twin-photon
instability requires the more challenging condition $4g^{2}/\kappa\Omega>1$.
Coherent effects associated with the artificial gauge field will be
observable in the strong coupling regime, $J_{\text{eff}}\gg\kappa$.
Parameter $J_{{\rm eff}}$ depends on the average phonon number $|\beta|^{2}$
in the classical mechanical oscillations, $J_{{\rm eff}}=g_{0}|\beta|\epsilon$
where $\epsilon$ is a small parameter which ensures the validity
of the perturbative treatment. When the mechanics is driven close
to its resonance we have $\beta\sim g_{0}\bar{n}/\Gamma$, where $\bar{n}$
is the photon number in the pump mode. We arrive essentially at the
condition $g^{2}\epsilon/\Gamma\kappa>1$ .

We note that observation of the optomechanical band structure would
not require ground state cooling (the effective temperature of the
system would only determine the relative weight of spectral features
observable in the emission spectrum discussed above or other quantities).

Other promising experimental platforms for optomechanical metamaterials
include microdiscs \cite{SM_Barclay2006,SM_Ding2010,SM_Zhang2012Sync} and
microtoroids \cite{SM_Armani2002,SM_Armani2007,SM_Verhagen2012} on a microchip
(which could be coupled via evanescent optical fields \cite{SM_Zhang2012Sync})
and superconducting circuits comprising microwave cavities parametrically
coupled to vibrational modes. Recent experimental demonstrations of
the latter systems employed micro- and nanomechanical beams \cite{SM_Regal2008,SM_Rocheleau2010}
and membranes \cite{SM_Teufel2011} as mechanical elements. Prospects
for both, mechanical and ``optical'' coupling in multi-mode systems
are currently investigated \cite{SM_Houck2012}.

\section*{Effects of disorder}

We briefly discuss the effects of static disorder introduced during
the fabrication of the optomechanical crystal structures. 

Mathematically speaking, the wave functions of a single particle hopping
on a disordered lattice are Anderson-localized both in 1D and 2D structures
for any finite disorder strength. This would be the fate of the eigenstates
of photons or phonons, respectively, and also of their optomechanically
hybridized versions, in the absence of dissipation and nonlinearities.
However, due to the finite decay rates for photons and phonons, as
well as the finite extent of the arrays, we have to be more quantitative
and estimate the localization length. 

A good first approximation to the actual situation is provided by
the Anderson model \cite{SM_1958_Andersonlocalization} of disorder
in a tight-binding lattice, with on-site potentials fluctuating randomly.
In this model, the potential values (i.e. optical or mechanical mode
frequencies for our case) are distributed evenly in the interval $[-W/2,+W/2]$,
where $W$ denotes the disorder strength. Note that the qualitative
results do not depend on the precise details of the assumed distribution.
If we take the hopping matrix element to be $J$, which in our case
stands for either photon or phonon hopping, the following results
can be extracted from the extended literature on strong localization
of electrons: In a 1D disordered chain, perturbation theory predicts
the localization length (measured in sites) for a wave at frequency
$\omega$ to be $24(4J^{2}-\omega^{2})/W^{2}$, i.e. maximal at the
band center, here assumed to be at $\omega=0$ \cite{SM_1993_Kramerreviewlocalization}.
 In 2D, numerical calculations \cite{SM_1981_MackinnonPRLandersonloc}
indicate that the localization length reaches about 100 sites already
for $W/J=5$, and several hundred sites for $W/J=4$. In conclusion,
both for 1D and especially for 2D the localization length will be
larger than the sample size for feasible optomechanical arrays, unless
the fluctuations of the local optical (or mechanical) mode frequencies
reach values on the order of the hopping matrix element itself. Therefore,
strong localization effects should not show up in the experiments
envisaged here. It might potentially be possible to observe the first
precursors of strong localization (diffusion of waves and weak localization,
or coherent backscattering for transport of photons through a slightly
disordered optomechanical array), but we have to leave the analysis
of these effects to future studies.

Current experiments indicate that the fluctuations of the on-site
optical frequencies are at most about 10\% of the hopping strength
\cite{SM_2008_notomilargephotonarrays}, i.e. well below the regime where
any strong effects of disorder would be present.

\end{document}